# 3D instability of a toroidal flow in the liquid partially covered by a solid film

V A Demin[1], M I Petukhov[1], A I Shmyrova[2]

[1]Perm State National Research University, Bukireva St. 15, 614068, Perm, Russia
[2]Institute of Continuous Media Mechanics, Ak. Koroleva St. 1, 614000, Perm, Russia

E-mail: demin@psu.ru

**Abstract.** Flow structure stability of a steady radial thermocapillary flow from the local heat source in cylindrical geometry has been studied numerically. The up boundary of the liquid was partially covered by the stationary film of an insoluble surfactant, pushed to the wall of the cavity. The calculations are fulfilled on the base of interfacial hydrodynamics equations by using mathematical package "Comsol Multiphysics". It is shown that the effect of a radial flow symmetry breakdown is appeared as a result of convective instability, so that the initial poloidal rotation of the liquid transforms into the azimuthal plane. The system of inclined steady volumetric vortices is formed under the film of surfactant. The inclination angle of the plane of rotation in dependence on the intensity of radial thermocapillary flow is studied. It is shown that the azimuthal vorticity can be regulated by the power of point-heat source and by free surface area. The increase of heating intensity leads to the growth of the azimuthal vorticity. Thus, the initially radially symmetric toroidal flow is divided into two parts with the origination of vortices in the azimuthal plane under the film.

## 1. Introduction

The flows, caused by the thermocapillary effect, are investigated theoretically and experimentally for a long time. The multiscale character of this phenomenon and numerous practical applications lead to the increasing interest at this field. At the date we can find a great number of numerical [1], analytical [2, 3] and experimental [4] researches of thermocapillary effect because it is appeared in different branches of physics, chemistry and biophysics. As a rule, the solution of these problems is complicated by mixing of the scientific approaches.

Interfacial effects are inseparably linked with the possible presence of surface-active substances on a free surface of the liquid. It leads to the significant change of the surface tension and consequent appearance of a definite response of a physical system. One of the most actual problems consists in the description and analysis of thermocapillary flow interaction with the surfactant.

The original problem, which motivated us for the present work, was published in papers of Homsy with co-authors [5, 6], where the behavior of the surfactant film on a non-uniformly heated gas–liquid interface, cojointly with volumetric convection, were investigated theoretically. The flow in a shallow two-dimensional slot was driven by a thermally induced surface stress in a fluid. It was demonstrated that for situations in which the adsorbed molecules are insufficient to result in a fully covered surface the interface is either clean, and subjected to non-zero stress, or contaminated by surfactant and satisfy to no-slip conditions. Also, Homsy introduced a non-dimensional governing parameter, the elasticity number, to describe the flow under the covered surface. The specific mathematical technique was used to obtain an analytical expression for the stream function in the vicinity of the leading edge of the stagnant zone in the limiting case of a creeping flow. These theoretical results laid the foundation for the understanding of the joint dynamics of a mutually interactive surfactant and liquid under an interface.

Experimental verification of these theoretical data was conducted in [7, 8]. The interface was heated from above with the help of a contactless optical system and open Hele–Shaw cell was used as the cavity. Contrary to the results stated in [6], these experiments gave qualitative difference in position of stagnant point. The essential difference in comparison with the problem [5, 6] was determined by three-dimensional character of the convective flow in the volume, another spatial distribution of the temperature on the interface and in the volume of the hydrodynamic system and additional property of compressibility of surfactant film. These specific features of the hydrodynamic system were taken into account in the theoretical model, presented in [8].

One more complexity of such problems can be connected with a possibility of convective flow, interacted with the surfactant film, to perform a three-dimensional motion. This rigorous deduction can be gotten on the basis of the analysis of experimental data [9]. The working cavity had a cylindrical form with open upper boundary. The heating was realized with the help of the light source and contactless optical system concentrating the heat flux in the geometrical center of the surface, covered initially by the uniform insoluble surfactant. Thermocapillary force induces the radial spreading of the liquid that leads to the redistribution of surfactant and consequent purification of interface in the middle of the cavity. Radial stagnant zone established on the surface periphery on this stage. It can be registered that the structure of the flow is much more complicated than in [8]. Namely, the initially symmetric radial motion changes the plane of rotation after it flow under the quasi-solid film and azimuthal vorticity appears near periphery of the cylindrical cuvette. The intensity of radial flow is determined by the power of the heat source. The bigger temperature difference on the surface causes more remote position of stagnant line from the center and the lower size of azimuthal vortices under the film.

The problem of such symmetry breakdown of the initially stable toroidal flow is very interesting from theoretical point of view. This article is devoted to the numerical investigation of the evolution of initially radial thermocapillary convection during the flow under the surfactant film.

## 2. Statement of the problem

Let us consider a cylindrical cavity with radius $R$ and height $h$, filled with the water which has free upper surface (Figure 1). As it was showed in [8], the transfer of liquid nearby the insoluble surfactant film is creeping: velocity of thermocapillary flow decreased by 3-4 orders of magnitude.

Thus, one can suppose, that in the frames of rough approximation the surfactant film interacting with flow can be considered as solid film which covers the part of the interface in the steady state. The area of free part of the liquid is determined by its radius $R^*$. The lateral boundary as well as the bottom of the cavity is rigid too. The interface of the working liquid is heated non-uniformly from above by a point source.

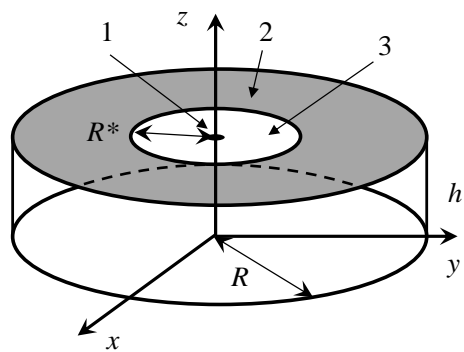

**Figure 1.** Geometry of the problem and coordinate system: 1 – point-heat source, 2 – solid film, 3 – free surface.

The heat- and mass transfer as in a volume as on the interface can be described by the following system of equations in Boussinesq approximation [10]:

$$\frac{\partial \boldsymbol{v}}{\partial t} + (\boldsymbol{v}\nabla)\boldsymbol{v} = -\frac{1}{\rho}\nabla p + v\Delta\boldsymbol{v} + g\beta T\boldsymbol{\gamma}, \quad (1)$$

$$\frac{\partial T}{\partial t} + (\boldsymbol{v}\nabla)T = \chi\Delta T, \quad (2)$$

$$\mathrm{div}\,\boldsymbol{v} = 0, \quad (3)$$

where $\boldsymbol{v}$, $p$, $T$ are the dimensional fields of velocity, pressure and temperature, respectively; $v$, $\chi$ are the coefficients of kinematic viscosity and thermal diffusivity, $\rho$ is the density of the fluid, $\beta$ is the coefficient of thermal expansion, $g$ is the magnitude of gravity acceleration, $\gamma$ is the unit vector, directed vertically upward. In contrast to general statement of the problem in the frames of interfacial hydrodynamics [11-13] the film of surfactant is motionless and specifies only by the parameter $R^*$ in present statement. Therefore, there is no necessity to describe the behaviour of surfactant by additional equation.

Temperature and velocity on the rigid boundaries are determined by the following conditions:

$$r = R: \quad T = 0, \boldsymbol{v} = 0, \qquad z = 0, h: \quad T = 0, \boldsymbol{v} = 0, \quad (4)$$

where $r$ is the radial coordinate. The conditions for temperature on the free boundary follow from the equation for the heat flux and correspond to the heating of the free surface in geometrical center:

$$z = h, 0 < r < R^*: \quad \kappa\frac{\partial T}{\partial z} = \frac{A}{2}\left(1 - erf\left(\frac{r^2}{k_1 R^2}\right)\right)\left(1 + erf\left(\frac{t - t_0}{k_2}\right)\right). \quad (5)$$

$A$ and $\kappa$ are the characteristic heat flux and thermal conductivity. Parameters $k_1$, $k_2$ and $t_0$ determine the radius of heating zone and the time for which the heating is established and get maximum intensity for given parameter $A$.

In the current statement, just like in previous one [7-9], devoted to the dynamics of an insoluble surfactant, the characteristic temperature difference is relatively small that allows to use the Gibbs isotherm in numerical simulation. So, the surface tension of the pure liquid $\sigma$ depends on the temperature in the standard way

$$\sigma = \sigma_0 - \sigma_T T.$$

Thus, the balance of tangential stresses on the free boundary is:

$$z = h, 0 < r < R^*: \quad \eta\frac{\partial v_x}{\partial z} = \frac{\partial \sigma}{\partial x} = -\sigma_T\frac{\partial T}{\partial x}, \quad \eta\frac{\partial v_y}{\partial z} = \frac{\partial \sigma}{\partial y} = -\sigma_T\frac{\partial T}{\partial y}, \quad (6)$$

where $\eta$ is the dynamic viscosity.

## 3. Numerical procedure

The equations system (1) – (6) was solved numerically using the mathematical software package "Comsol Multiphysics" and computing cluster "Triton" of Institute of Continuous Media Mechanics of UB RAS. A non-uniform tetrahedral mesh was generated in the calculations with crowding near the center of the interface (Figure 2). For chosen number of cells, we obtained fast and well-converged numerical results. In the numerical simulation we used the MUMPS software application (where MUMPS stands for multifrontal massively parallel sparse direct solver).

The condition for stability of the calculation scheme is the smallness of the Courant number: Co = $\tau|U|/H$, where $\tau$ and $H$ are the steps by time and coordinate, respectively; and $U$ is the value of rate. In physical terms, the Courant condition means that a liquid particle should not move for a time step by more than one spatial step. The Courant number is computed for all cells of the calculation region at each instant time.

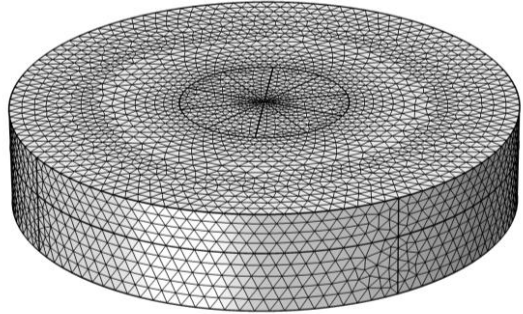

**Figure 2**. The scheme of the working mesh.

The size of the cylindrical cavity is $R = 0.05$ m, $h = 0.02$ m. The parameters of working liquid are close to the properties of water: $\nu = 10^{-6}$ m²/s, $\chi = 10^{-8}$ m²/s, $\sigma_T = 10^{-4}$ N/mK, $\beta = 10^{-4}$ 1/K, $\kappa = 0.6$ Wt/mK, $\rho = 10^3$ kg/m³. In the process of numerical simulation parameters of heating in boundary condition (5) had following values: $k_1 = 0.04$, $k_2 = 180$ s, $t_0 = 300$ s.

## 4. Results of calculation

The most natural flow for this convective system is the radial-symmetric spreading caused either by thermocapillary effect, if the part of the surface is open, or by gravitational convection, if whole surface is closed by a surfactant. These flows can be observed in numerical simulation (Figure 3).

These results also show that thermocapillary flow, established on the free surface, shifts the centres of the vortices to the solid wall of the cavity (at the left side on the Figure 4(a)). At the same time flow in the closed cavity have almost symmetric form in poloidal section (Figure 4(b)). If motionless surfactant partially covers the surface, so that $0 < R^* < R$, the initially radial spreading can be strongly affected after the flow under the film. Laminar creeping flow of the liquid near the sudden obstacle causes the notable rearranging of velocity components to satisfy mass conservation law for incompressible liquid.

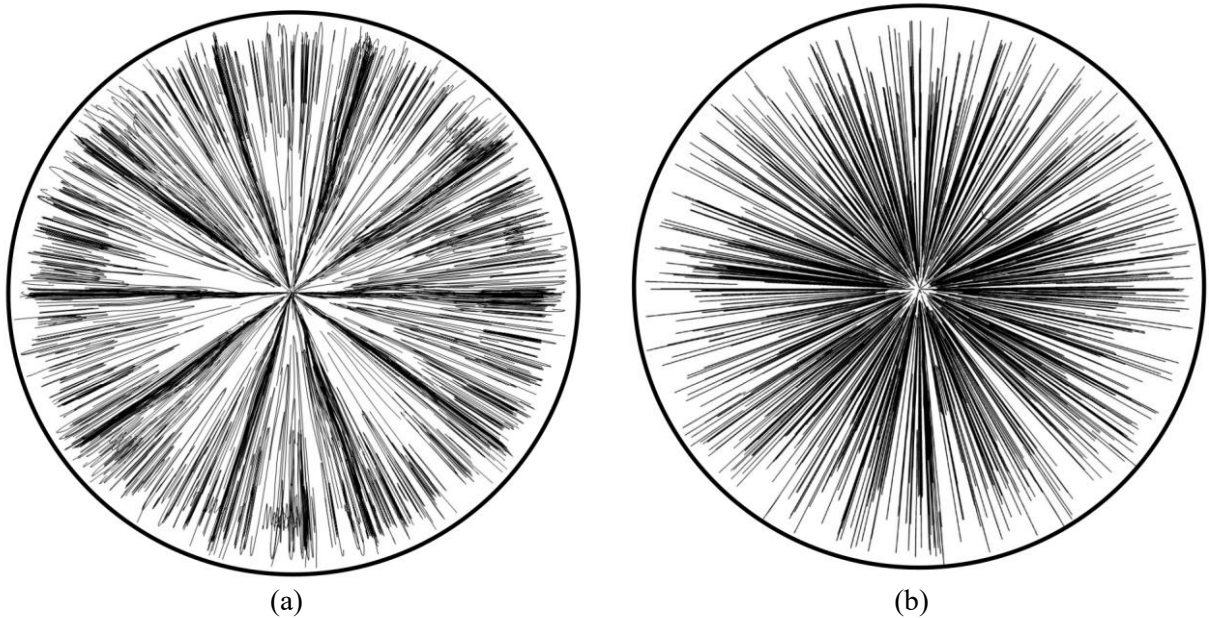

(a)          (b)

**Figure 3**. Flows in cylindrical cavity for completely opened surface in the case $R^* = R$, $A = 5 \cdot 10^{-4}$ Wt/m² (a), and closed surface for $R^* = 0$, $A = 10 \cdot 10^{-4}$ Wt/m² (b). View from above.

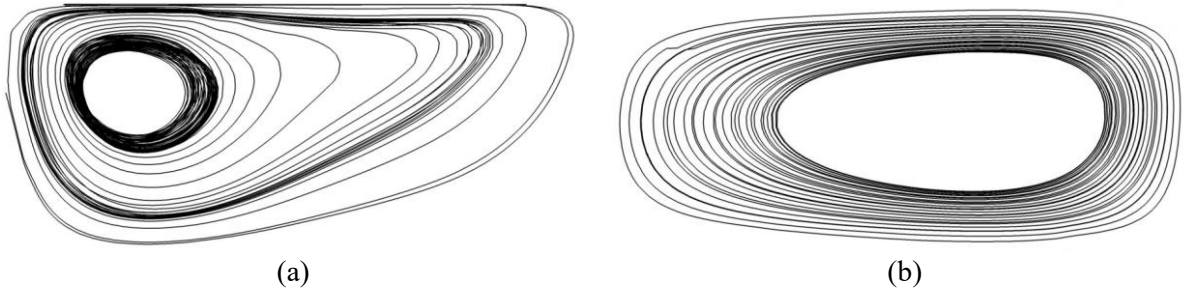

(a)                (b)

**Figure 4**. Projections of flows for completely opened surface (a): $R^* = R$, $A = 5 \cdot 10^{-4}$ Wt/m$^2$), and closed surface (b): $R^* = 0$, $A = 10 \cdot 10^{-4}$ Wt/m$^2$. Side view.

One can say that transformation of this flow is connected with instability in a diffuser which is expressed in an appearing of azimuthal component of velocity. Only under the film in our case, the liquid particles obtain corresponding horizontal components of velocity and reorganize vortices into azimuthal plane (Figure 5).

One can see that mostly this flow still consists of radially oriented trajectories, following to the thermocapillary spreading out of the center to the periphery. In their turn, azimuthal-oriented vortices, appeared under the film, have secondary nature. After displacement of stagnant zone, whose role plays the boundary of the rigid film the vortices originate in compliance with the definite value of wave number. The bigger radius of open surface corresponds to the bigger wave number of the vortices (Figure 5).

Numerical simulation has showed that such rearranging of flow appears only for relatively notable intensity of heating. If the heating is weak the vortices will be toroidal despite the presence of rigid plane. It means that there must be continuous transfer of vortices under the film from vertical orientation to horizontal one when the heat intensity increases. This transfer is described by the Figure 6. It is seen that the increasing of area covered by surfactant leads to the appearance of conditions for which vortices with notable inclination angle are formed for weaker heating.

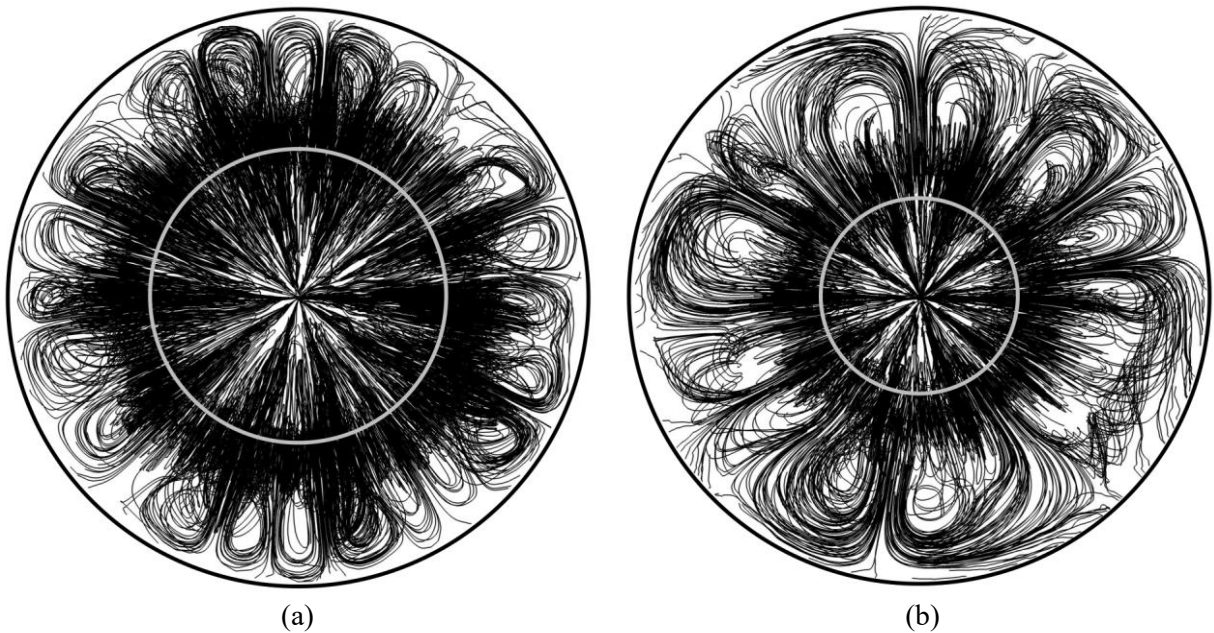

(a)                (b)

**Figure 5**. Flows in cylindrical cavity for partially opened surface (a): $R^* = R/2$, $A = 24 \cdot 10^{-4}$ Wt/m$^2$, and closed surface (b): $R^* = R/3$, $A = 15 \cdot 10^{-4}$ Wt/m$^2$. View from above. Gray line corresponds to stagnant line.

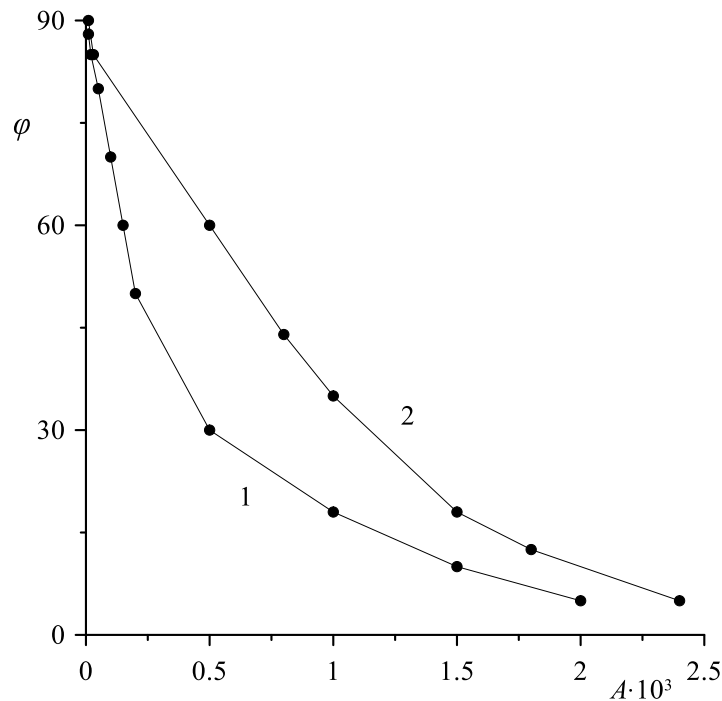

**Figure 6.** Inclination of vortices in respect of vertical axis. 1 – $R^* = R/2$, 2 – $R^* = R/3$.

It is also obvious, that vortices with azimuthal component of velocity cannot be completely horizontal. The trajectories have complex structure of descending turning lines, where nonzero vertical component of velocity always exist. Therefore, the graphs in Figure 6 can be only directed towards zero.

## 5. Conclusion
The behavior of radial-symmetric incident flow on the quasi-solid film of stationary surfactant has been considered numerically. The surfactant is approximated as rigid plane, partially covered free surface of the liquid. Under the film the flow loses initial radial symmetry and redirects its motion from downward to the azimuthal plane. This leads to establishing of corresponding vortices under the film in the background of radial flow. The nature of such effect can be connected with instability, appearing in the diffuser. We could find this type of instability for incompressible fluid in the cavity with infinite radius. In our system this instability is induced by the presence of stagnant line on the surface. On the one hand this approach is primary and movability of the film must be included in the model, if we want to describe real surface-active matter. On the other hand considered model describes hydrodynamic system with maximal friction on the walls. Conditions of the vortices appearance with azimuthal component of velocity are only complicated by this fact. The region of azimuthal instability of the flow could be wider for the systems with movable surfactant.

**Acknowledgements**
This work was supported by the Russian Science Foundation (Grant No. 19-71-00097).